\documentclass[12pt,preprint]{aastex}

\def\alwaysmath#1{\ifmmode{#1}\else{$#1$}\fi}

\def\arcsec{\hbox{$^{\prime\prime}$}}

\def\compulb{COM~J1701$-$3006B~}

\newcommand\CHANDRA{{\it Chandra }}

\def\ltsima{$\; \buildrel < \over \sim \;$}

\def\gtsima{$\; \buildrel > \over \sim \;$}

\def\lsim{\lower.5ex\hbox{\ltsima}} 
\def\gsim{\lower.5ex\hbox{\gtsima}} 
\def\lapp{\ifmmode\stackrel{<}{_{\sim}}\else$\stackrel{<}{_{\sim}}$\fi} 
\def\gapp{\ifmmode\stackrel{>}{_{\sim}}\else$\stackrel{<}{_{\sim}}$\fi} 
 
\slugcomment{Submitted to ApJ Letters} 
 
\shorttitle{The optical companion to PSRJ1701-3006B in NGC\,6266}  
\shortauthors{Cocozza et al.} 
 
\begin{document} 
 
\title{A puzzling millisecond pulsar companion in 
NGC\,6266\altaffilmark{1}} 
 
\author{ 
G. Cocozza\altaffilmark{2,3},  
F. R. Ferraro\altaffilmark{2}, 
A. Possenti\altaffilmark{4}, 
G. Beccari\altaffilmark{2}, 
B. Lanzoni\altaffilmark{2}, 
S. Ranson\altaffilmark{5},  
R. T. Rood\altaffilmark{6}, 
N. D'Amico\altaffilmark{4,7}} 

\altaffiltext{1}{Based on observations with the NASA/ESA HST, 
obtained at the Space Telescope Science Institute, which is operated 
by AURA, Inc., under NASA contract NAS5-26555.} 
\altaffiltext{2}{Dipartimento di Astronomia Universit\`a 
di Bologna, via Ranzani 1, I--40127 Bologna, Italy,}  
\altaffiltext{3}{INAF, Osservatorio Astronomico di Bologna,  
via Ranzani 1, I--40127 Bologna, Italy} 
\altaffiltext{4}{INAF, Osservatorio Astronomico di Cagliari, 
Loc. Poggio dei Pini, Strada 54, I--09012 Capoterra, Italy} 
\altaffiltext{5}{National Astronomy Observatory, 520 Edgemont Road,  
Charlottesville, VA, 22903, USA} 
\altaffiltext{6}{Astronomy Department, University of Virginia, 
P.O. Box 400325, Charlottesville, VA, 22904, USA} 
\altaffiltext{7}{Dipartimento di Fisica Universit\`a di Cagliari, 
Cittadella Universitaria, I-09042 Monserrato, Italy} 
 
 
\begin{abstract} 
We report on the optical identification of the companion to the 
eclipsing millisecond pulsar PSR\,J1701$-$3006B in the globular 
cluster NGC\,6266. A relatively bright star with an anomalous red 
colour and an optical variability ($\sim$ 0.2 mag) that nicely 
correlates with the orbital period of the pulsar ($\sim$ 0.144 days) 
has been found nearly coincident with the pulsar nominal 
position. This star is also found to lie within the error box position 
of an X-ray source detected by Chandra observations, thus supporting 
the hypothesis that some interaction is occurring between the pulsar 
wind and the gas streaming off the companion. Although the shape of 
the optical light curve is suggestive of a tidally deformed star which 
has nearly completely filled its Roche lobe, the luminosity ($\sim 
1.9\,L_\odot$) and the surface temperature ($\sim 6000$\,K) of the 
star, deduced from the observed magnitude and colours, would imply a 
stellar radius significantly larger than the Roche lobe radius. 
Possible explanations for this apparent inconsistency are discussed. 
\end{abstract} 
 
  
\section{Introduction}  
\label{sec:intro} 
 
NGC\,6266 (M62) is one of the most massive and brightest ($M_V=-9.19$; 
Harris 1996\footnote{Everywhere in this paper, we refer to the  
data at {\tt http://www.physics.mcmaster.ca/Globular.html}}) 
Galactic Globular Clusters (GCs), also is characterized by high values of
the central density (log $\rho_0 \sim 5.47$, with $\rho_0$ in units of
$M_\odot\,$pc$^{-3}$; Beccari et al. 2006).  It displays a King-model
density profile with an extended core ($\sim 19\arcsec$) and a modest
value of the concentration parameter ($c=1.5$; Beccari et al. 2006).

Six binary millisecond pulsars (MSPs) have been discovered in M62
(D'Amico et al. 2001\nocite{damico01}; Jacoby et
al. 2002\nocite{jacoby02}; Possenti et al. 2003\nocite{P03}, hereafter
P03) and it ranks fifth of the GCs in wealth of MSPs, after Terzan 5,
47 Tucanae, M15 and M28\footnote{See {\tt
    http://www.naic.edu/{~}pfreire/GCpsr.html} for a catalog of MSPs
  in GCs with relative references}. P03 presented phase--connected
timing solutions, yielding precise celestial coordinates, for three of
the MSPs in NGC\,6266. One of them, PSR\,J1701$-$3006B (hereafter
PSR\,6266B) displays partial or total eclipses of the radio signal at
1.4 GHz near its superior conjunction (in the convection adopted
throughout this paper, this corresponds to orbital phase $\phi
=0.25$), clearly due to gas streaming off the companion. The pulsar
orbit is circular, with a projected semi-major axis of only $\sim
0.11\,R_\odot$.  P03 suggested two options for explaining the
behaviour of the PSR\,6266B system. The first option is that the
pulsar companion is a non-degenerate bloated star, whose mass loss is
sustained by ablation of its loosely bound surface layers by the
relativistic wind emitted by the pulsar. In this case, PSR\,6266B may
resemble PSRs B1957+20 (Fruchter et al. 1990\nocite{fbb+90}) and
J2051$-$0827 (Stappers et al. 2001\nocite{sbl+01}): the optical light
curve of their companion star presents a maximum when the side of the
companion facing the pulsar points toward the observer (i.e., at the
pulsar inferior conjunction: $\phi=0.75$). This is a clear signature
of the irradiation of the companion surface by the pulsar flux. The
second option has the pulsar companion as a tidally deformed star
overflowing its Roche lobe due to the internal nuclear evolution. In
this case, the system may be more like PSR\,J1740$-$5340 (D'Amico et
al. 2001\nocite{damico01}; Ferraro et al. 2001\nocite{ferraro01}),
where irradiation effects are negligible (Orosz \& van Kerkwijk
2003\nocite{ov03}) and the optical light curve of the companion is
dominated by ellipsoidal variations (Ferraro et
al. 2001\nocite{ferraro01}), i.e. it shows two maxima at quadratures
($\phi=0.0$ and $0.5$).
 
We present the optical identification of the companion to PSR\,6266B,
based on high-quality phase-resolved photometry obtained with the {\it
  Hubble Space Telescope} (HST), and X-ray emission detected with
\CHANDRA.
 
\section{Observations and data analysis} 
{\sc HST observations.}  The photometric data presented here consist
of a set of high-resolution images obtained on August 1 2004 by using
the Advanced Camera for Survey (ACS), on-board the HST, and retrieved
from the {\it ESO/ST-ECF Science Archive}.  The data comprise three
$R$-band exposures ($2\times340\,$s and $1\times30\,$s exposures), two
$B$-band exposures (of 120 and 340\,s) and four $H_\alpha$-band
exposures (of 340, 1050, 1125, 1095\,s).  The three longest exposures
in $H_\alpha$ are the combination of three sub-exposures.  An
additional set of Wide Field Planetary Camera 2 (WFPC2, Prop. 10845,
PI:Ferraro) data has been secured through the $H_\alpha$ filter with
the specific aim of testing the possible variability of the companion:
ten\footnote{Only 9 exposures have been used in the analysis since one
  was made useless by a cosmic ray passage.} 1200\,s exposures were
taken between May 2 and May 5 2007.  In order to best resolve stars in
the most crowded regions, the Planetary Camera (with a resolution of
$0.046\arcsec/$pixel) of the WFPC2 was pointed at the cluster center.
The photometric analysis of the ACS dataset has been carried out using
the ACS module of the Dolphot\footnote{The DOLPHOT packages, including
  the ACS module and all the documentation, can be found at {\tt
    http://purcell.as.arizona.edu/dolphot/}.} package \cite{dolphin},
using the FLT.fits datasets for the photometry, and the drizzled
images as references for the geometric distortion correction.  We set
the photometry parameters as recommended in the Dolphot manual,
obtaining a final catalog of $B$, $R$ and $H_\alpha$ magnitudes,
calibrated according to Sirianni et al. (2005).  The final catalog is
reported to an absolute astrometric system by cross-correlating the
stars in common with the data set of Beccari et al. (2006), which has
been astrometrized by using the standard stars from the new {\it Guide
  Star Catalog II} ($GSCII$) lying in the considered field of view
(FoV).  At the end of the procedure, the rms residuals were of the
order of $\sim 0.5\arcsec$ both in RA and Dec; we assume this value as
representative of the astrometric accuracy.
 
In order to carefully search for any variability from the MSP
companion, we reanalyzed the $H_\alpha$ images in a small area around
the nominal position of PSR\,6266B by using ROMAFOT (Buonanno et
al. 1983). This package has been specifically developed to perform
accurate photometry in crowded fields, and it also allows a visual
inspection of the quality of the PSF-fitting procedure.  In
particular, we extracted $50\times 50$-pixels ($\sim 2.5\arcsec \times
2.5\arcsec$) sub-images from the ten original $H_\alpha$ ACS frames,
each centered on the nominal position of PSR\,6266B.  In order to
optimize the detection of faint objects, we performed the search
procedure on the median image. Then, we adopted the resulting mask
with the star positions to the individual sub-images, and performed
the PSF-fitting procedure separately on each of them. We adopted the
same strategy for the analysis of the nine WFPC2 $H_\alpha$ images,
after the careful alignement with the previously astrometrized ACS
frames. The resulting instrumental magnitudes were transformed to a
common system, and we compiled a final catalog with coordinates and
magnitudes for all the stars identified in the considered 19
sub-images. The photometric calibration of the instrumental magnitudes
and the absolute celestial coordinates were determined by using all
the stars in common with the overall ACS catalog, calibrated and
astrometrized as discussed above.

{\sc Chandra Observations.}  NGC\,6266 was observed for 63 ks on May
2002 with the \CHANDRA Advanced CCD Imaging Spectrometer (ACIS). The
ACIS-S3 chip was pointed at the cluster center, its FoV ($8\arcmin
\times 8 \arcmin$) entirely covering the half-mass radius
($1.23\arcmin$; Harris 1996).  The data were reduced using the CIAO 
software (v. 3.3.0), reprocessing the level 1 event files; the task 
{\tt acis\_run\_hotpix} was used to generate a new badpixel file. Then 
we applied the corrections for pixels randomization, good time 
intervals and grades.  Our analysis includes only photons with ASCA 
grades of 0, 2, 3, 4 and 6. About 1.5 ks of observations were not 
considered in the analysis because of the high background level.  We 
searched for discrete sources in our level 2 event file using {\tt 
Wavdetect} (Freeman et al. 2002), performing the source detection 
separetely on the 0.2-1, 1-2, 2-8, and 0.2-8 keV bands. The detection 
threshold was set to $10^{-5}$, and the source detection was performed 
increasing the sequence of wavelets from 1 to 16 by factors of 
$\sqrt{2}$. More than 110 sources were detected in the 0.2-8 keV band 
in the entire chip.  The astrometry of all the detected sources was 
corrected by using the Aspect Calculator provided by the \CHANDRA 
X-Ray Center. Then we searched for possible optical counterparts both
in our HST images and catalog. About 5 bright stars and 2 faint and
extended objects (likely background galaxies) were found to coincide
with the X-ray sources well outside the half-mass radius, where the
stellar density is relatively low. We therefore used these optical
counterparts for extending the astrometric solution of HST to the
\CHANDRA sources, thus obtaining the same accuracy in the absolute 
astrometry for both the optical and the X-ray sources. 
 
\section{Results}  
\begin{figure}  
\plotone{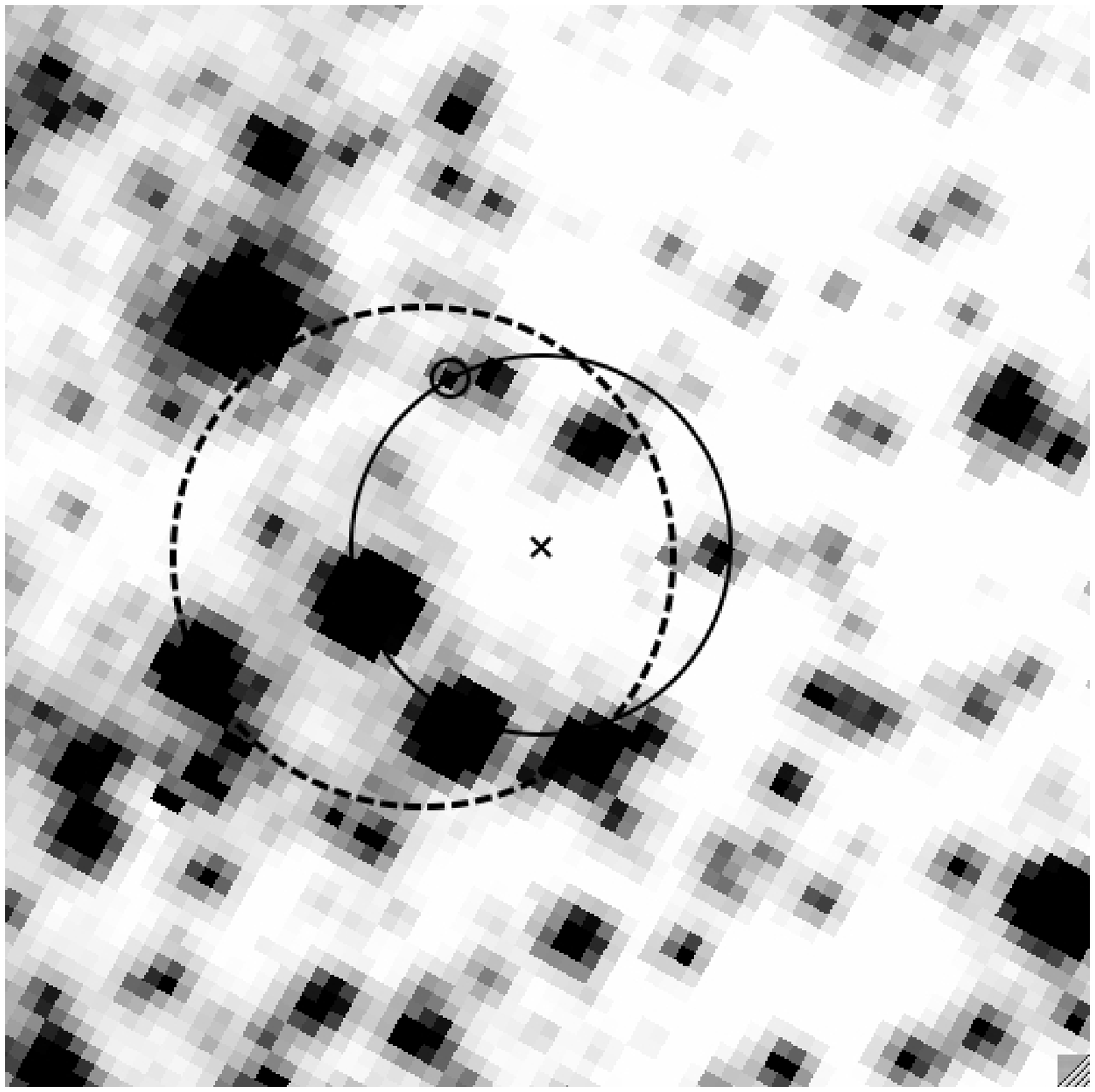}  
\caption{Finding chart for \compulb showing the median-combined WFPC2 
  $H_\alpha$ image. The region covers $3\arcsec \times 3\arcsec$; the 
  cross marks the nominal position of PSR\,6266B, with the $1\sigma$ 
  error circle ($0.5\arcsec$, solid line). The smaller circle shows 
  the optical counterpart, whereas the dashed line represents the 
  $1\sigma$ circle ($0.7\arcsec$) in the \CHANDRA position. North is 
  up and East is to the left. 
\label{fig:chart}}  
\end{figure}

Figure \ref{fig:chart} shows the $3\arcsec \times 3\arcsec$ WFPC2 
$H_\alpha$ map centered on the PSR\,6266B nominal position (marked 
with a {\it cross}), as derived from timing (see P03). An accurate 
photometric analysis has been carried out for all the $\sim30$ stars 
found in our catalog within a distance of $1.5\arcsec$ from the 
PSR\,6266B position. This correspond to 3 times the $1\sigma$ 
uncertainty ($0.5\arcsec$) in the absolute astrometry of the HST data. 
In particular, we have extracted 19 individual images from our dataset 
(10 exposures with ACS and 9 with WFPC2) and compared the resulting 
magnitudes for each object in order to search for variability and 
estimate the typical photometric uncertainties at different magnitude 
levels. Only one source in the catalog (hereafter named COM\,6266B, 
marked with a small circle in Fig.~\ref{fig:chart}) showed $H_\alpha$ 
variability ($\Delta H_\alpha \sim 0.2$ mag) significantly larger than 
the typical rms magnitude fluctuations of stars of similar luminosity 
($\delta H_\alpha \sim 0.02$ mag in ACS data, and $\delta H_\alpha 
\sim 0.04$ mag in WFPC2 data).  Its celestial coordinates are 
$\alpha_{\rm J2000}=17^{\rm h}01^{\rm m} 12^{\rm s}\!.690$ and 
$\delta_{\rm J2000}=-30\arcdeg 06\arcmin 48\farcs61$, whereas the mean 
apparent magnitudes are $B=20.58$, $V=19.48$, $R=19.02$, 
$H_\alpha=18.65$ ($V$ is from Beccari et 
al. 2006\nocite{Beccari06}). It is located at $0.5''$ from PSR\,6266B 
just at the edge of its positional error circle. 
 
Figure \ref{fig:CMDs} reports the $(R,\,H\alpha-R)$ and $(R,\,B-R)$ 
Color Magnitude Diagrams (CMDs) for all the stars detected in the 
$20\arcsec\times20\arcsec$ box centered on the PSR\,6266B nominal 
position. COM\,6266B is marked with a {\it large filled triangle} in 
both the CMDs. As apparent from the figure, the star has almost the 
same luminosity of the cluster Turn Off, but it shows an anomalous red 
color which locates it out of the Main Sequence. The photometric 
properties imply that it is not a degenerate star and that it has a 
moderate $H_\alpha$ excess, indicating the presence of ionized gas 
surrounding the system. These findings are consistent with the 
irregulaties seen in the radio signal of PSR\,6266B and supports the 
scenario that COM\,6266B is a mass losing star. 
 
In order to firmly assess the physical connection between COM\,6266B
and the binary pulsar PSR\,6266B, we have accurately investigated the
time scale of the optical variability\footnote{Note that COM\,6266B
  has a close companion located at only 4 pixels ($\sim 0\farcs19$)
  West in the ACS map (see Fig.\ref{fig:chart}). This object turns out
  to be a normal Sub-Giant Branch star, slightly brighter ($\sim0.5$
  mag in $B$, $R$ and $H_\alpha$) than COM\,6266B and not displaying
  any evidence of variability.  Although very close to each other, the
  two stars are clearly separable in the high resolution ACS and WFPC2
  images and we have verified that the photometric analysis of
  COM\,6266B is not affected by the presence of this close star.}  (in
the 0-1 day range). The $H_\alpha$ time series (19 points) was
processed using the GRATIS $\chi^2$ fitting routine, a code developed
at the Bologna Astronomical Observatory in order to study the
periodicity of variable stars (see, e.g., Clementini et
al. 2000\nocite{clem+00}). Following the procedure fully described in
Ferraro et al. (2001), the most significant (99\% confidence
interval)\footnote{The significance level has been obtained from
  the reduced $\chi^2$ of the Fourier time series fit (at different
  modulation periods) to the $H_\alpha$ data (see for example Figure
  2 of Ferraro et al. 2001).}  periodicity was found at a period of
$0.1446\pm 0.0015$ day, consistent, within the uncertainties, with the
orbital period of PSR\,6266B derived from timing
($P_b=0.1445454303(6)$ day, where the figure in parenthesis is the
uncertainty, at 99\% confidence level, on the last quoted digit;
P03). Given that, we have folded the 19 magnitude values, by using
$P_b$ and the reference epoch of the PSR\,6266B radio ephemeris
($T_0=52047.258199(2)$; P03). The results are shown in
Fig.\ref{fig:lc}. As apparent, the ACS and the WFPC2 data (obtained
about 3 years later) agree with each other, drawing a well defined
light curve with two distinct maxima at about $\phi=0.0$ and
$\phi=0.5.$ This fact, as well as the phases and the amplitudes of the
minima, confirms that the optical modulation is associated with the
pulsar binary motion and strongly suggests that COM\,6266B is a
deformed star overflowing its Roche lobe (see \S 1).

Additional properties of this system can be derived from the analysis
of the \CHANDRA data. We found an X--ray source located at only $\sim
0\farcs4$ from COM\,6266B and at $0\farcs3$ from the nominal radio
position of PSR\,6266B. The derived coordinates of the X-ray source
are $\alpha_{\rm J2000}=17^{\rm h}01^{\rm m}12^{\rm s}\!.700$ and
$\delta_{\rm J2000}=-30\arcdeg 06\arcmin 49\farcs08,$ and the
$1\sigma$ uncertainty area on its position is encircled in
Fig.\ref{fig:chart} (dashed line). Given the number of sources (50)
detected within the half-mass radius of NGC\,6266, the probability of
a chance superposition of a X--ray source within $0\farcs3$ from the
radio position of PSR\,6266B is $\sim 7\times10^{-4},$ strongly
supporting the association of the detected source with the PSR\,6266B
system.  The (background substracted) photons counts are
$18.3_{-4.2}^{+5.3}$ (soft: 0.2--1 keV), $23.7_{-4.9}^{+6.0}$ (medium:
1--2 keV) and $9.8_{-3.1}^{+4.3}$ (hard: 2--8 keV), implying
HR1$\simeq 1.3$ and HR2$\simeq 0.4$ (where HR1=medium/soft counts and
HR2=hard/medium counts). Hence the source is harder than the typical
MSP population observed in GCs (e.g. Grindlay et
al. 2002\nocite{grind02}) and the HR1, HR2 values resemble those of a
source in which the pulsar wind shocks the material released from the
companion (see, e.g., the cases of PSR\,J1740$-$5340 in NGC\,6397, and
PSR\,J0024$-$7204W in 47Tuc; Grindlay et al. 2002\nocite{grind02}, and
Bogdanov et al. 2005\nocite{bogd05}, respectively). Even if the small
number of photons prevents a detailed spectral analysis, for an
absorption column density $N_H=2.2\times 10^{21}\,{\rm cm^{-2}}$
(Pooley et al. 2003\nocite{Pooley03}), the counts in each band and the
HR1 and HR2 values are consistent\footnote{The PIMMS tool has been
  used for these estimates: {\tt
    http://heasarc.gsfc.nasa.gov/Tools/w3pimms.html}} with a power law
model having a photon index $\simeq 2.5$, and a total unabsorbed flux
$F_X\sim 10^{-14}\,{\rm erg\ cm^{-2} s^{-1}}$ in the 0.2-8 keV band,
translating (for a distance of 6.6\,kpc; Beccari et
al. 2006\nocite{Beccari06}) into a X--ray luminosity of $L_X\sim
6\times 10^{31}\,{\rm erg\ s^{-1}}$.

\section{Discussion}

Many observed features of COM\,6266B (namely, its anomalous red
colour, the $H_\alpha$ excess, the shape of the light curve, the X-ray
emission from the binary) are suggestive of a tidally deformed star
which is experiencing heavy mass loss, similar to the system found in
NGC\,6397 by Ferraro et al. (2001\nocite{ferraro01}). On the other
hand, other photometric properties of the source do not fit this
picture, as described below.

\begin{figure}  
\plotone{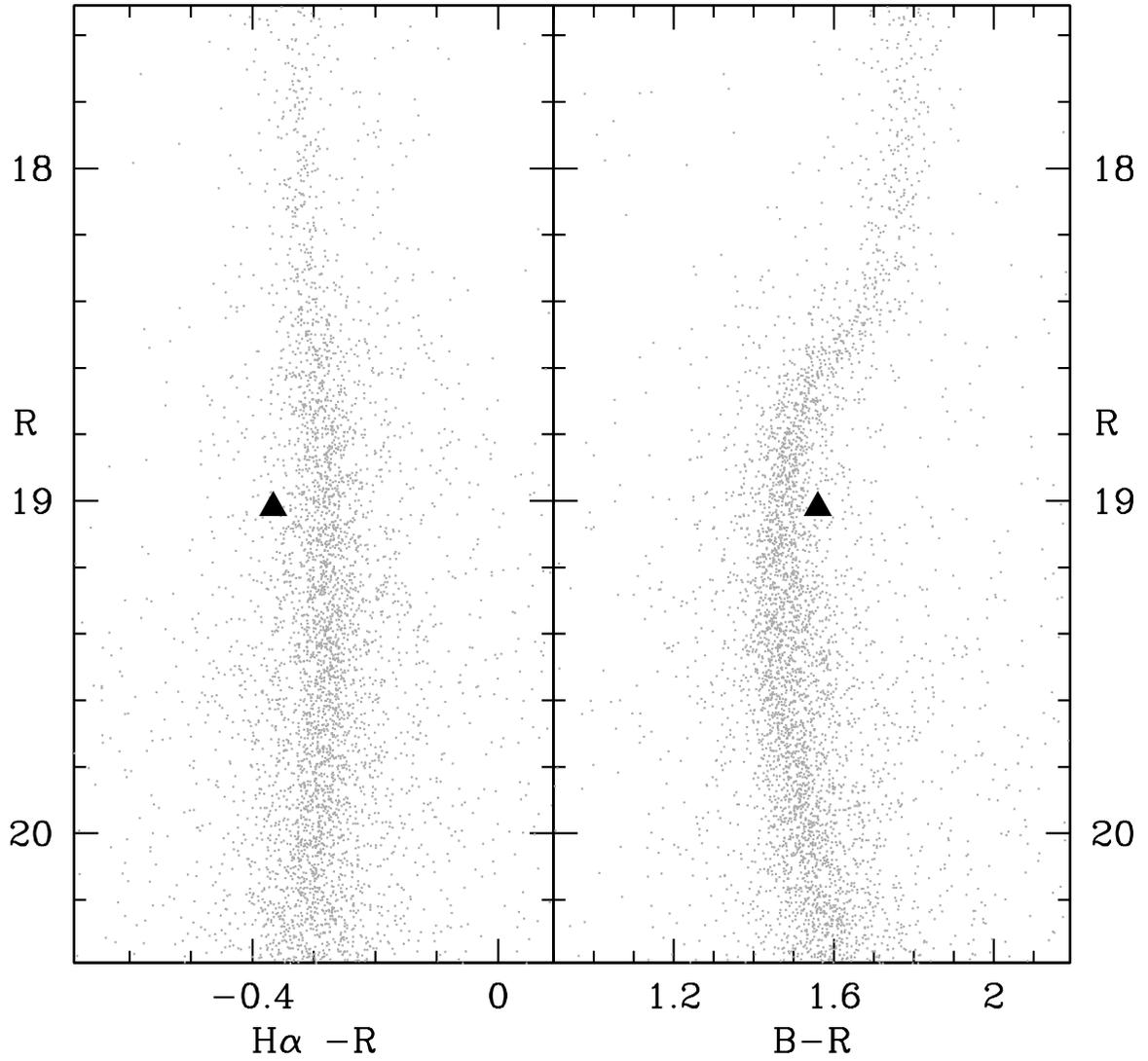} 
\caption{ $(R,\,H_\alpha-R)$ and $(R,\,B-R)$ CMDs for stars identified 
  in a region of $20\arcsec \times20\arcsec$ centered at the nominal 
  position of PSR\,6266B. The optical counterpart to the pulsar 
  companion COM\,6266B is marked with a {\it large filled triangle}.
\label{fig:CMDs}} 
\end{figure}

\begin{figure} 
\plotone{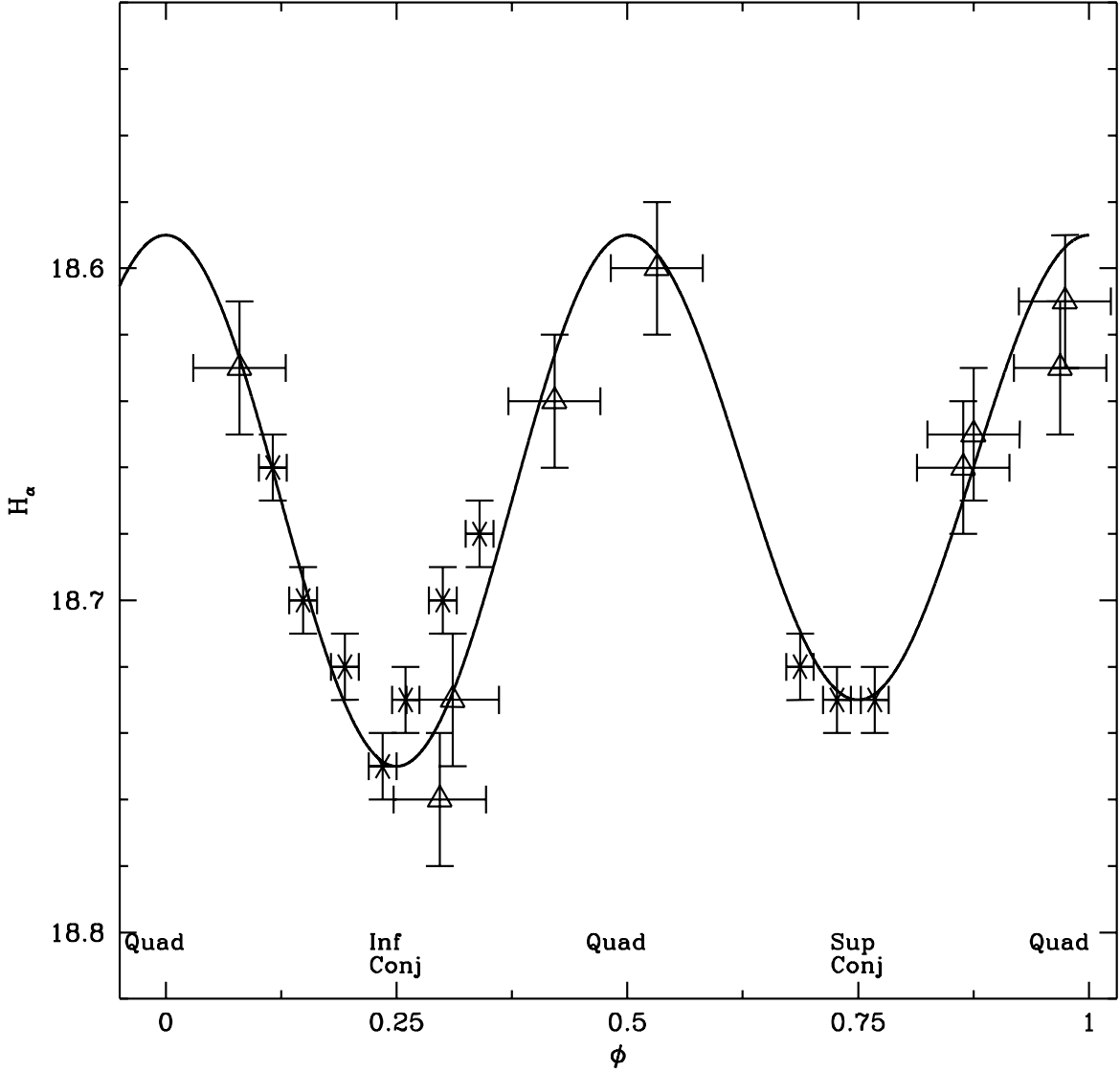} 
\caption{ $H_\alpha$ light curve of COM\,6266B obtained by using the 
  period and the reference epoch of the radio ephemeris of 
  PSR\,6266B. Asterisks represent ACS archive observations performed 
  in August 2004; large empty triangles are the WFPC2 data collected 
  in May 2007. The phases of quadrature and conjunction of COM\,6266B 
  are reported for clarity. The {\it solid line} represents the 
  Fourier time series (trucated to the second harmonic) that best fits 
  the data. 
\label{fig:lc}}  
\end{figure}

By assuming the reddening $E(B-V)\sim 0.47$ and the distance quoted by
Beccari et al. (2006), and adopting $[Fe/H]=-1.1$ and an age of 12 Gyr
for NGC\,6266, we have determined the physical parameters of
COM\,6266B from the comparison of its position on the CMDs
(Fig.\,\ref{fig:CMDs}) with the isochrones of Pietrinferni et
al. (2004\nocite{Pietr04}). The resulting effective temperature,
bolometric luminosity and radius of this star are $T_{\rm eff}\sim
6000\pm 500\,$K, $L_{\rm bol}\sim 1.9\pm0.2\,L_\odot,$ and $R\sim
1.2\pm 0.2\,R_\odot$, respectively, where the quoted uncertainties are
conservative estimates, essentially due to the uncertanties in the
distance modulus ($\pm 0.15$ mag) and in the reddening ($\pm 0.05$
mag).  If the luminosity variations shown in Fig.~\ref{fig:lc} are
mainly due to tidal deformations of the companion star, the stellar
radius $R$ is expected to be of the order of the Roche lobe radius
$R_L$. The latter can be estimated from the mass function and the
projected semi-major axis of PSR\,6266B (derived from pulsar timing;
P03), provided that a pulsar mass $M_{\rm psr}$ and an orbital
inclination $i$ are assumed. As already discussed in P03, the presence
of radio eclipses indicates that $i$ is not small.
Conservatively taking $i\gsim20\arcdeg$ and assuming $M_{\rm
  psr}=1.4\,M_\odot$, the companion star is expected to span a mass
interval ranging from $0.125\,M_\odot$ ($i=90\arcdeg$) to
$0.41\,M_\odot$ ($i=20\arcdeg$)\footnote{Note that in both cases the
  orbital separation of the system turns out to be $a\sim
  1.4\,R_\odot$.}, corresponding to $R_L\sim 0.26\,R_\odot$ and
$0.40\, R_\odot$, respectively. The value of $R_L$ is only slightly
affected by the choice of $M_{\rm psr},$ being, for instance,
$R_L\lsim 0.45\, R_\odot$ for $0.7\, M_\odot\le M_{\rm psr}\le 3.0\,
M_\odot$ (a safely large neutron star mass range; see, e.g., Freire et
al. 2008\nocite{freire08}).  Note that in order to match the estimated
$R_L$ with the observed $R$ it should be $i\sim 3\arcdeg$ (for $M_{\rm
  psr}=1.4\, M_\odot$), thus implying an unreliable companion mass of
$\sim 6\, M_\odot$. In summary, for all plausible binary parameters,
it turns out that the observed stellar radius of COM\,6266B is
significantly larger ($R\gsim 3\,R_L$) than the expected Roche lobe
radius of the companion to PSR\,6266B, i.e. COM\,6266B appears $\gsim
10$ times brighter than the expected luminosity of any pulsar
companion having an effective temperature $T_e=6000~K.$
 
Under the hypothesis that COM\,6266B is the companion to PSR\,6266B, 
what is the origin of this remarkable discrepancy? A number of 
possibilities are examined below.

{\it (i)} The binary system could not belong to the cluster. In this 
case the adopted distance would be inappropriate and the value of 
the radius deduced from the luminosity would be largely overestimated. 
On the other hand, the celestial position of PSR\,6266B is at less 
than $2\arcsec$ from the cluster center, and the pulsar shows a 
negative value of the spin period derivative, as expected if 
PSR\,6266B lies in the cluster potential well. 
 
{\it (ii)} The optical luminosity of COM\,6266B could be dominated by
strong non-thermal processes, possibly triggered by the pulsar spin
down power $L_{\rm sd}\lsim 10 \,L_\odot$ (see P03).  Also in this
case the value of the stellar radius deduced from the observed
luminosity could be significantly overestimated.  However, At the
orbital separation of $1.4 R_\odot$, the energy captured by a Roche
lobe filling companion would be of the order of $L\sim 0.1-0.2
L_\odot$ which is too low to explain the observed value. Moreover, it
is hard to explain how a large non-thermal component may generate a
light curve with maxima that fall at orbital phases nicely consistent
with those expected for light curves modulated by ellipsoidal
variations.

{\it (iii)} COM\,6266B could be the blend of two low-luminosity stars:
one star would be the optically variable companion to PSR\,6266B and
the other one a non-variable star in the foreground
(background). While stellar blends are not uncommon in the highly
crowded GC cores, this possibility only partially alleviates the
problem. In fact, by assuming that COM\,6266B is the blend of two
stars of equal mean $H_\alpha$ magnitude (i.e., $H_\alpha=19.35$), the
observed ellipsoidal modulation could only be produced if the pulsar
companion varies by $\Delta{H_\alpha}\sim 0.35$ mag, which is a rather
extreme value for variabilities induced by tidal
distortions. Moreover, even in this case, the star would have a radius
$R \gsim 1.7 R_L$. Matching $R$ with $R_L$ would require that
COM\,6266B is e.g. the blend of a non-variable star with
$H_\alpha=18.7$ and a star with mean magnitude $H_\alpha \sim 21$
varying by $\Delta{H\alpha}>3 $ mag. We note that similar
considerations would apply also if PSR\,6266B were a triple system.

In conclusion, while option {\it (iii)} seems to be a quite unlike
{\it ad hoc} scenario, options {\it (i)} and {\it (ii)} will be
finally addressed as soon as spectroscopic data are available for the
system. This will allow us to investigate the nature (thermal or not
thermal) of the optical emission and the association of the source to
the cluster (via the determination of the radial velocity), hence
shedding some light on the nature of this puzzling object.

\acknowledgements This research was supported by the Agenzia Spaziale 
Italiana (ASI I/088/06/0), {\it PRIN-INAF2006} and by the Ministero 
dell'Istruzione, dell'Universit\`a e della Ricerca, and it is part of 
the {\it Progetti Strategici di Ateneo 2006} granted by University of 
Bologna. AP and NDA also acknowledge financial support from {\it 
PRIN-MIUR2005}. We thank the Referee for the careful reading of the 
manuscript.

\clearpage
 
\clearpage  
 
\end{document}